\documentclass[aps,prl,twocolumn,superscriptaddress]{revtex4-1}

\usepackage{graphicx}
\usepackage{bm}
\usepackage{rotating}
\usepackage{amssymb,amsmath}
\usepackage{multirow}
\usepackage{color}
\usepackage{hyperref}
 
\begin{document}

\title{CaIrO$_3$: a Spin-Orbit Mott Insulator Beyond the  $j_{\mathrm{eff}}=1/2$
Ground State}

\author{M.~Moretti~Sala}
\email{marco.moretti@esrf.fr}
\affiliation{European Synchrotron Radiation Facility, BP 220, F-38043 Grenoble
Cedex, France}

\author{K.~Ohgushi}
\affiliation{Institute for Solid State Physics, University of Tokyo, Kashiwa,
Chiba 277-8581, Japan}

\author{A.~Al-Zein}
\affiliation{European Synchrotron Radiation Facility, BP 220, F-38043 Grenoble
Cedex, France}

\author{Y.~Hirata}
\affiliation{Institute for Solid State Physics, University of Tokyo, Kashiwa,
Chiba 277-8581, Japan}

\author{G.~Monaco}
\affiliation{Dipartimento di Fisica, Universit\`a di Trento, via Sommarive 14,
38123 Povo (TN), Italy}
\affiliation{European Synchrotron Radiation Facility, BP 220, F-38043 Grenoble
Cedex, France}

\author{M.~Krisch}
\affiliation{European Synchrotron Radiation Facility, BP 220, F-38043 Grenoble
Cedex, France}

\begin{abstract}

In CaIrO$_3$ electronic correlation, spin-orbit coupling, and tetragonal
crystal field splitting are predicted to be of comparable strength. However, the
nature of its ground state is still object of debate, with contradictory
experimental and theoretical results. We probe the ground state of CaIrO$_3$
and assess the effective tetragonal crystal field splitting and
spin-orbit coupling at play in this system by means of resonant inelastic x-ray
scattering. We conclude that insulating CaIrO$_3$ is \emph{not} a
$j_{\mathrm{eff}}=1/2$ iridate and discuss the consequences of our finding to
the interpretation of previous experiments. In particular, we clarify how the
Mott insulating state in iridates can be readily extended beyond the
$j_{\mathrm{eff}}=1/2$ ground state.

\end{abstract}

\maketitle

Spin-orbit coupling is the main ingredient for 5$d$ transition metal oxides
to form novel electronic states of matter, such as the recently
discovered Mott insulating state in
Sr$_2$IrO$_4$~\cite{Kim2008,Moon2008,Kim2009}. This insulating
behaviour is unexpected in iridates perovskites as, for a half-filled shell with
spatially extended orbitals, electronic correlation was thought to be
negligible. Instead, the role of electronic correlation is enhanced here by
spin-orbit coupling, through the formation of the so-called
$j_{\mathrm{eff}}=1/2$ ground state. Its realization arises from the
interaction of strong spin-orbit coupling ($\zeta\sim0.5$ eV) and cubic crystal
field (10Dq $\sim 3$ eV), and is perturbed by short- and long-ranged
anisotropies which could cause departures from the $j_{\mathrm{eff}}=1/2$ ground
state~\cite{Katukuri2012,Zhang2013,Boseggia2013a,
MorettiSala2013b_arxiv}. A small, but sizable tetragonal
contribution $\left| \Delta \right| \sim0.01$ eV to the cubic crystal field 10Dq
was detected in both Sr$_2$IrO$_4$~\cite{Boseggia2013a} and
Ba$_2$IrO$_4$~\cite{MorettiSala2013b_arxiv}, with opposite signs, in
agreement with recent theoretical calculations~\cite{Zhang2013}.
In these cases, however, $\left| \Delta \right| \ll \zeta \ll$ 10Dq and the
$j_{\mathrm{eff}}=1/2$ ground state in Sr$_2$IrO$_4$ and Ba$_2$IrO$_4$ is not in
doubt. Structural distortions are instead more pronounced in the insulating
post-perovskite CaIrO$_3$~\cite{Hirai2009}; thus the cubic symmetry of the
crystal field is expected to be drastically lowered and the
$j_{\mathrm{eff}}=1/2$ ground state to be severely altered. The robustness of
the $j_{\mathrm{eff}}=1/2$ ground state against structural distortions, in
particular octahedral rotations and elongations, as well as chemical
substitution has been mostly tested by means of resonant X-ray magnetic
scattering (RMXS), on the basis of the nearly vanishing intensity at the L$_2$
absorption edge~\cite{Kim2009}. Following this criterion, a number of
``$j_{\mathrm{eff}}=1/2$ iridates'' have been
identified~\cite{Boseggia2012,Boseggia2012a,JWKim2012,Boseggia2013,
Calder2013_arXiv}, including CaIrO$_3$~\cite{Ohgushi2013}. However,
this interpretation has been widely
controversed~\cite{Chapon2011,Haskel2012,MorettiSala2014}. Indeed, a
unified picture has not been reached yet: the interpretation of RMXS
results~\cite{Ohgushi2013}, was supported by LDA+SO+U~\cite{Subedi2012}, but
contradicted by \emph{ab-initio} quantum chemistry~\cite{Bogdanov2012}
calculations. The latter predict a large splitting of the $t_{2g}$ states that
give rise to a strongly unbalanced occupation of the $xy$, $yz$ and $zx$
orbitals, while for the $j_{\mathrm{eff}}=1/2$ ground state the three orbitals
contribute with an equal weight of 1/3. Remarkably, despite contradicting
evidences about the details of the electronic ground state, consensus is reached
on the magnetic interactions, with a strong antiferromagnetic coupling along the
$c$-axis and a weak ferromagnetic one along the $a$-axis, which stabilize canted
long range antiferromagnetism~\cite{Ohgushi2006} and characterize CaIrO$_3$ as a
quasi-one-dimensional antiferromagnet~\cite{Bogdanov2012}. 

In this Letter, we use resonant inelastic X-ray scattering (RIXS) at the
Ir L$_3$ edge to solve the puzzle of the ground state in CaIrO$_3$. RIXS is a
powerful technique for the study of correlated electron systems mostly devoted
to the study of high-$T_c$ superconducting and insulating
cuprates in the past years~\cite{Ament2011}. Recently, RIXS was applied
to the investigation of magnetic excitations in correlated iridium
oxides. Following the initial theoretical suggestion of Ament \emph{et
al.}~\cite{Ament2011a}, it was demonstrated that magnon dispersion
could be studied in Sr$_2$IrO$_4$~\cite{Kim2012,Lupascu2013_arxiv} and
Sr$_3$Ir$_2$O$_7$~\cite{JKim2012}. In the present work, we focus on the
inelastic response of CaIrO$_3$ in the energy range relevant to spin-orbital
excitations~\cite{Kim2012,Liu2012,Hozoi2012,Gretarsson2013} and we determine 
the effective tetragonal crystal field and spin-orbit
coupling acting on $t_{2g}$ levels by comparing the results to a single-ion
model~\cite{Ament2011a,Hozoi2012,MorettiSala2014}. We
obtain $\zeta=0.52$ and $\Delta=-0.71$ eV and
therefore conclude that the departure of CaIrO$_3$ from the
$j_{\mathrm{eff}}=1/2$ state is unambigous, in agreement with
\emph{ab-initio} calculations~\cite{Bogdanov2012}.

RIXS measurements were performed at the new inelastic X-ray
scattering beamline of ESRF (ID20-UPBL06). Two different set-ups were used in
order to optimize the flux and energy-resolution of the beamline. The incident
radiation was monochromatized by a high heat-load double crystal Si(111)
monochromator and a post-monochromator for further bandwidth reduction. In
the high energy-resolution mode, a Si(844) back-scattering channel-cut
reduced the incident photon bandwidth down to 15 meV at 11.215
keV~\cite{MorettiSala2013}; in the low energy-resolution set-up, a Si(311)
channel-cut monochromator provided a bandwidth of about 350 meV. A
Kirkpatrick-Baez mirror system focused the X-rays to a spot size of 10$\times$20
(V$\times$H) $\mu$m$^2$ at the sample position. The scattered X-rays were
energy-analyzed by a Rowland-type spectrometer, equipped with 5 diced Si(844)
analyzers ($R=2$ m), and detected by a 5-element Maxipix detector, with a pixel
size of 55$\times$55 $\mu$m$^2$~\cite{Ponchut2011}. The overall
energy-resolution was about 25 (350) meV in the high (low) energy-resolution
mode. A single crystal of CaIrO$_3$ was grown by the flux method as in
Ref.~\onlinecite{Ohgushi2013}. CaIrO$_3$ has a post-perovskite structure (space
group $Cmcm$~\cite{Sugahara2008}), composed of edge-sharing (corner-sharing)
IrO$_6$ octahedra along the $a$-axis ($c$-axis), in which each octahedron is
compressed along the corner-sharing O direction (the local $z$ direction) with a
bond length ratio of 0.97. Because of the alternating rotation of the octahedra
around the $a$-axis, the local $z$- and the crystallographic $c$-axis form an
angle of $\pm23^\circ$, thus forming a zig-zag chain of Ir-O-Ir bonds. The
lattice parameters are $a=3.147$, $b=9.859$ and $c=7.290$ \AA~\cite{Hirai2009}.
CaIrO$_3$ is an insulator and undergoes a transition to a canted antiferromagnet
at $T_N=115$ K~\cite{Ohgushi2006}, in which the strong spin-orbit coupling
stabilizes a striped-type magnetic order (magnetic moments are aligned parallely
along $a$ and mostly antiparallely along $c$, with a small canting in the
direction of the $b$-axis)~\cite{Ohgushi2013}. Throughout the experiment, the
sample was kept at a temperature of 15 K.

\begin{figure}
	\centering
		\includegraphics[width=.9\columnwidth]{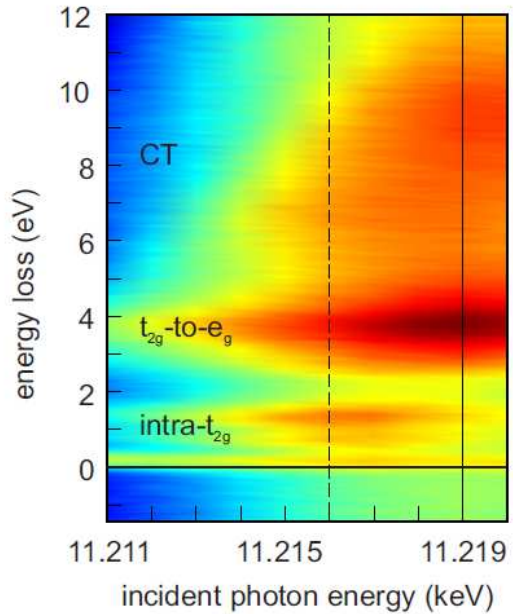}
\caption{RIXS color map of CaIrO$_3$. The continuous vertical line corresponds
to the maximum of the absorption profile at 11.219 keV. The dashed vertical
line, instead, corresponds to the incident photon energy used for the high
energy-resolution RIXS spectrum of Fig.~\ref{fig2} (11.216 keV).}\label{fig1}
\end{figure}

Fig.~\ref{fig1} shows a low energy-resolution RIXS map of CaIrO$_3$. In
RIXS, the incident photon energy is tuned to an absorption edge, the $L_3$ in
our case, corresponding to the transition of an electron from the 2p$_{3/2}$ to
the 5d states, thus creating a deep core hole. The system then relaxes to a less
excited final state by filling the core hole with the same or another electron,
possibly creating a low energy excitation. The incident energy is scanned
across the L$_3$ absorption line and spectra are recorded up to 12 eV energy
loss. An elastic line and magnetic excitations are found close to the zero
energy loss line. At increasing energy losses, we assign features to
intra-$t_{2g}$ ($t_{2g}^5\rightarrow t_{2g}^5$), $t_{2g}$-to-$e_g$ 
($t_{2g}^5\rightarrow t_{2g}^4e_g^1$) and charge-transfer (CT) excitations
(as indicated in the figure), following previous RIXS
studies~\cite{Ishii2011}. In this work, we concentrate on
intra-$t_{2g}$ excitations only. Their intensity (including that of the
elastic line and magnetic excitations) is enhanced for incident photon energies
about 3 eV below the main absorption peak. This is a common feature of
several iridium oxides, which was observed in both RMXS and RIXS
experiments~\cite{Boseggia2012,Liu2012}. It can be understood by
considering the electronic structure of an Ir$^{4+}$ ion in a large cubic
crystal field: with 5 electrons filling the 5d states, $e_g$ states are empty
and one hole is left in the $t_{2g}$ states. The maximum of the absorption
line (11.219 keV) corresponds to the transition of an electron into the 5d $e_g$
state, while the contribution of the $t_{2g}$ states is minor, as it scales with
the number of unoccupied final states. On the other hand, it is reasonable to
assume that intra-$t_{2g}$ excitations are more effectively probed when a
2p$_{3/2}$ electron is directly promoted in the 5d $t_{2g}$ state, i.e. for
incident photon energies tuned at $\sim$ 10Dq below the main absorption line
(11.216 keV). 

\begin{figure}
	\centering
		\includegraphics[width=\columnwidth]{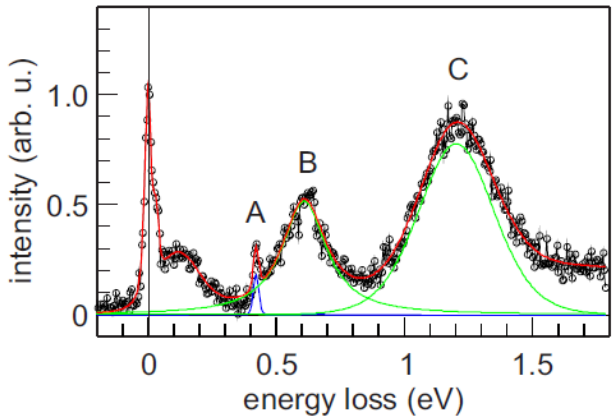}
\caption{RIXS spectrum of CaIrO$_3$ at a selected momentum transfer,
$\mathbf{Q}=(1,2,10)$ r.l.u.. Continuous lines are fit to the experimental
data.}\label{fig2}
\end{figure}

For the high resolution measurements, we fixed the incident photon energy at
11.216 keV to enhance $t_{2g}$ excitations. A representative spectrum is shown
in Fig.~\ref{fig2}. The low energy range comprises the elastic line and a
broad structure of magnetic origin, not discussed here. The 0.3-1.6 eV energy
range is dominated by two intense broad features (B and C) and a weak,
energy-resolution limited peak (A), similar to the excitation spectrum of
Na$_2$IrO$_3$ and Li$_2$IrO$_3$~\cite{Gretarsson2013,Gretarsson2013a}. The
assignment of features B and C to local excitations across crystal
field split states~\cite{Gretarsson2013} was initially
debated~\cite{Foyevtsova2013}, but then supported by recent
calculations~\cite{BHKim2013_arxiv}. We therefore assign features B and C of
Fig.~\ref{fig2} in analogy to the case of Na$_2$IrO$_3$ and Li$_2$IrO$_3$. The
assignment of features B and C is further strengthened by \emph{ab-initio}
calculations for CaIrO$_3$ predicting a splitting of the $j_{\mathrm{eff}}=3/2$
states by 0.6-0.7 eV~\cite{Bogdanov2012}, in agreement with the energy
difference of about 0.6 eV for features B and C found in our experiment.

In the following we discuss the spin-orbit excitations (features B and C in
Fig.~\ref{fig2}) in more detail. Although of mostly local origin, it should  be
mentioned that they show a weak, but non negligible dispersion versus momentum
transfer, in the order of 0.05 eV, indicating that they retain some non-local
character. In the present work, however, we aim at determining the size of the
effective tetragonal crystal field splitting and spin-orbit coupling in
CaIrO$_3$ and the fine details of the band structure will be neglected. In
fact, their bandwidth is one order of magnitude smaller than the energy at
which these excitations occur, and therefore inter-site interactions can be
considered as small perturbations to the local crystal field. The
\emph{effective} parameters $\Delta$ and $\zeta$ then include any kind of
renormalization due to non-local effects. A dedicated study of the
transferred momentum dependence of such excitations is, however, desirable.

\begin{figure}
	\centering
		\includegraphics[width=\columnwidth]{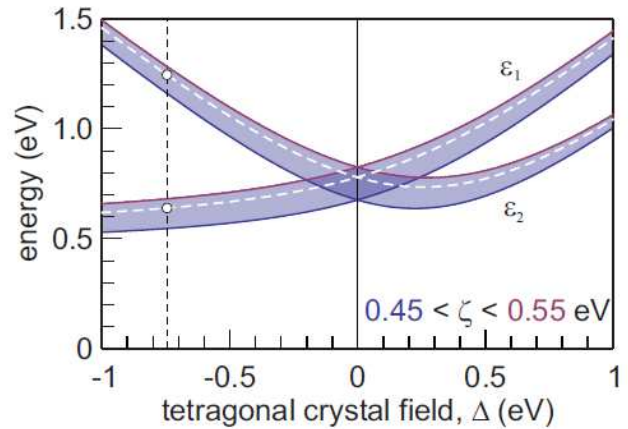}
\caption{Energy diagram of spin-orbit excitations calculated
according to Eqs.~(\ref{E1}) and ~(\ref{E2}) as a function of
$\Delta$ for $0.45<\zeta<0.55$ eV. White dashed lines
correspond to the estimated value of $\zeta=0.52$ eV, while the
white dots correspond to the actual value of the spin-orbit excitations
in CaIrO$_3$.}\label{fig3}
\end{figure}

The spectra were fitted to three Pearson functions: feature A turns out
to be energy-resolution limited, at an energy of about 0.42 eV, very similar to
the value reported for Na$_2$IrO$_3$ and Li$_2$IrO$_3$~\cite{Gretarsson2013}.
Features B and C are centered at 0.66 and 1.22 eV, respectively, which
corresponds to the largest intra-$t_{2g}$ splitting ever reported in iridium
oxides. In order to better understand the nature of these excitations, and to
assess the effective tetragonal crystal field splitting and spin-orbit coupling
in CaIrO$_3$, we adopt a single-ion
model~\cite{Ament2011a,Liu2012,Ohgushi2013,Hozoi2012,MorettiSala2014,
MorettiSala2013b_arxiv}. The weak momentum transfer dependence of these
excitations suggests a dominant intra-site character, for which a local model is
justified. Since 10Dq is sufficiently large (10Dq $\sim 3$ eV), 5$d$ $e_g$
states are neglected and the interacting Hamiltonian for one \emph{hole} in
the 5$d$ $t_{2g}$ states is then written as
\begin{equation}\label{hamilt}
 \mathcal{H} = \zeta \mathbf{L}\cdot\mathbf{S}-\Delta  L_z^2,
\end{equation}
in which tetragonal crystal field splitting ($\Delta<0$ for compressed
octahedra) and spin-orbit coupling are treated
on equal footing. For $\Delta=0$, the ground state is represented by the
$\left|j_{\mathrm{eff}}=1/2\right\rangle$ doublet and the
excited states by the $\left|j_{\mathrm{eff}}=3/2\right\rangle$ quadruplet.
For arbitrary values of $\zeta$ and $\Delta$, the eigenstates of $\mathcal{H}$
are three non-degenerate Kramers doublets, which we generically label
$|0,\pm\rangle$, $|1,\pm\rangle$ and $|2,\pm\rangle$ ($|0,\pm\rangle$
being the ground state wave function). The corresponding eigenvalues, $E_0$,
$E_1$ and $E_2$, are reported, for example, in
Ref.~\onlinecite{MorettiSala2014}. Here we are interested in transitions
from the $|0,\pm\rangle$ ground state to the $|1,\pm\rangle$ and $|2,\pm\rangle$
excited states. The corresponding energies, relative to the ground state, at
which the RIXS transitions occur, are given by
\begin{eqnarray}
 \varepsilon_1=E_1-E_0 &=& \frac{\zeta}{4} \left[3+\delta +\sqrt{9 +
\delta(\delta-2) }\right] \label{E1} \\
 \varepsilon_2=E_2-E_0 &=& \frac{\zeta}{2} \sqrt{9 + \delta(\delta-2)},
\label{E2}
\end{eqnarray}
where $\delta=2\Delta/\zeta$. These are reported in Fig.~\ref{fig3} as a
function of $\Delta$ for $0.45<\zeta<0.55$ eV. For $\Delta=0$,
$\varepsilon_1=\varepsilon_2=3\zeta/2$ as the $|1,\pm\rangle$ and
$|2,\pm\rangle$ states merge into the $\left|j_{\mathrm{eff}}=3/2\right\rangle$
quadruplet when the $j_{\mathrm{eff}}=1/2$ ground state is realized. In
general, however, the energy of the two RIXS excitations is a function of
both $\Delta$ and $\zeta$. 

In our case, the only solution with physical meaning is found by imposing
$\varepsilon_1=0.66$ and $\varepsilon_2=1.22$ eV (white dots in the diagram of
Fig.~\ref{fig3}), from which one obtains $\zeta=0.52$ and $\Delta=-0.71$ eV. The
value of $\zeta$ compares favourably with previous estimates in other iridium
oxides~\cite{Liu2012,Gretarsson2013,Boseggia2013,MorettiSala2013b_arxiv}, and
in particular to the recently calculated value of 0.47 eV by Bogdanov \emph{et
al.}~\cite{Bogdanov2012} for the specific case of CaIrO$_3$. The sign of
$\Delta$ is consistent with structural studies reporting a compression
of the IrO$_6$ octahedra~\cite{Hirai2009}, while its magnitude implies a minor
contribution of the $xy$ orbital to the ground state wave funtion (only
10\%), in agreement with \emph{ab-initio} quantum chemistry calculations which
predict a splitting of $t_{2g}$ states in the absence of spin-orbit coupling in
the order of 0.63-0.76~\cite{Bogdanov2012}. Noteworthy, $\left|\Delta\right|$ is
more than one order of magnitude larger than that of the prototypical
$j_{\mathrm{eff}}=1/2$ compounds, Sr$_2$IrO$_4$ ($\Delta=-0.01$
eV)~\cite{Boseggia2013a} and Ba$_2$IrO$_4$ ($\Delta=0.05$
eV)~\cite{MorettiSala2014}. Therefore, $\left|\Delta\right| > \zeta$ and we can
safely state that CaIrO$_3$ is \emph{not} a $j_{\mathrm{eff}}=1/2$ iridate, with
a dominant $\left(|yz,\mp\rangle\pm\imath|zx,\mp\rangle\right)/\sqrt{2}$
character of the ground state wave-function. 

We now discuss the size and direction of the magnetic moments in comparison
to existing experimental results and the sign of magnetic interactions. The size
of the magnetic moment is a function of both $\zeta$ and
$\Delta$~\cite{MorettiSala2014}: with the above determined values, the
expectation value for the $z$ component of the magnetic moment turns out to be
1.7 $\mu_B$ in CaIrO$_3$, i.e. larger than the magnetic moment of 1 $\mu_B$
expected for a perfectly localized $j_{\mathrm{eff}}=1/2$ state. Following the
interpretation of magnetization data of Ref.~\onlinecite{Ohgushi2013}, we
conclude that the magnetic moments in CaIrO$_3$ are canted along the $b$-axis,
with a canting angle of $\sim2^\circ$, i.e. about $21^\circ$ away from the local
$z$-axis of the IrO$_6$ octahedra. It should be noted here that the discrepancy
between RMXS and RIXS results is reconciled in view of their non-trivial
dependence on magnetic moment direction~\cite{MorettiSala2014}. Indeed,
considering the actual value of $\Delta$ and $\zeta$, and the direction of the
magnetic moments, one obtains a Ir $L_2$/$L_3$ RMXS intensity ratio of
0.1\%, which is below the detectability limit of 0.3\% reported in the
experiment~\cite{Ohgushi2013} (the calculated Ir $L_2$/$L_3$
RMXS intensity ratio would be 24\% in the case of magnetic moments aligned along
the local $z$-axis). Magnetic interactions giving rise to the stripe-type canted
antiferromagnetism of CaIrO$_3$ discussed above were explained in the framework
of an ideal $j_{\mathrm{eff}}=1/2$ state~\cite{Ohgushi2013}, for which
theoretical arguments~\cite{Jackeli2009} predict antiferromagnetic
(ferromagnetic) coupling along the corner-shared (edge-shared) bonds.
CaIrO$_3$ seems to obey these rules. However, as long as the occupancy of the
$yz$ and $zx$ orbitals is identical and their phase relation preserved, the same
theoretical arguments apply and the sign of the magnetic interactions remains
unchanged. Indeed, the fact that the coupling is ferromagnetic along $a$ and
antiferromagnetic along $b$, despite the severe departure of CaIrO$_3$ from the
$j_{\mathrm{eff}}=1/2$ ground state, is also supported by quantum chemistry
calculations~\cite{Bogdanov2012}.

\begin{figure}
	\centering
		\includegraphics[width=\columnwidth]{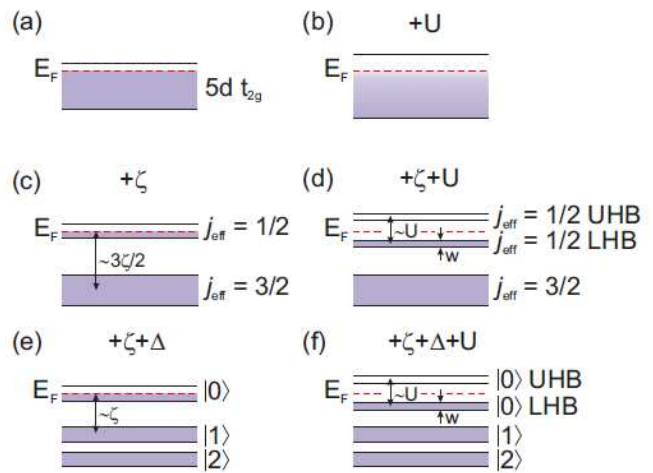}
\caption{Schematic representations of the band structure in
iridates with 5$d^5$ configurations in the absence of perturbations (a),
with spin-orbit coupling (c) and spin-orbit coupling plus tetragonal crystal
field splitting (e). Panels (b), (d) and (f) correspond to panels (a), (c) and
(e), respectively, when the Hubbard term $U$ is taken into account.}\label{fig4}
\end{figure}

We are now in the position to discuss the transport properties
of iridates and their connection to the $j_{\mathrm{eff}}=1/2$ ground state.
The latter was originally invoked to explain the insulating behaviour of
Sr$_2$IrO$_4$ and readily extended to other ``$j_{\mathrm{eff}}=1/2$''
compounds. In Fig.~\ref{fig4} we explain and extend the concept of a spin-orbit
Mott insulator beyond the specific case of ``$j_{\mathrm{eff}}=1/2$'' iridates.
One has to consider the 5$d$ $t_{2g}$ states, whose bandwidth
in the absence of perturbations would be too large for a reasonable Hubbard
energy $U$ to open a gap; rather, the density of states at the Fermi energy
would be only slightly reduced, as in Fig.~\ref{fig4}(b). Crucially, spin-orbit
coupling in the absence of a tetragonal crystal field splits the otherwise
degenerate $t_{2g}$ states and a half filled $j_{\mathrm{eff}}=1/2$ bands is
isolated, with a much reduced bandwidth ($w$) compared to the original one. As
$U>w$, lower (LHB) and upper (UHB) Hubbard bands are created, thus turning the
system into an insulator (Fig.~\ref{fig4}(d))~\cite{Kim2008}. In the case
of CaIrO$_3$, however, the large tetragonal crystal field degrades the
$\left|j_{\mathrm{eff}}=1/2\right\rangle$ ground state into the generic
$|0\rangle$. Nevertheless, the smallest splitting between the
$j_{\mathrm{eff}}=1/2$- and $j_{\mathrm{eff}}=3/2$-derived bands is
$\zeta$ ($\varepsilon_1$ in the limit $\Delta\rightarrow-\infty$), i.e. only a
factor 3/2 smaller than that in the pure $j_{\mathrm{eff}}=1/2$ ground state.
Therefore, for $U>w$ LHB and UHB bands are formed and the system retains its
insulating character, although the ground state wave function differs
significantly from $\left|j_{\mathrm{eff}}=1/2\right\rangle$ (indeed,
$|0,\pm\rangle=\pm|xy,\pm\rangle$ for $\Delta\rightarrow+\infty$ and
$|0,\pm\rangle=\left(|yz,\mp\rangle\pm\imath|zx,\mp\rangle\right)/\sqrt{2}$ for
$\Delta\rightarrow-\infty$). A rough estimate of $U$ in CaIrO$_3$ can be naively
extracted by adopting the interpretation, though
debated~\cite{Foyevtsova2013,BHKim2013_arxiv}, of feature A in Fig.~\ref{fig2}
as the excitation across the Mott gap~\cite{Gretarsson2013}: we obtain
$U\simeq0.4$ eV which is consistent with the band gap of 0.34 eV
deduced from resistivity measurements~\cite{Ohgushi2006,Subedi2012} and places
CaIrO$_3$ in the scenario of Fig.~\ref{fig4}(f).

In conclusion, i) we solve the controversy concerning the ground state
of CaIrO$_3$~\cite{Ohgushi2013,Bogdanov2012}: CaIrO$_3$ is \emph{not} a
$j_{\mathrm{eff}}=1/2$ iridate, due to the large tetragonal crystal field.
Indeed, we estimate the effective tetragonal crystal field splitting and
spin-orbit coupling to be $\Delta=-0.71$ and $\zeta=0.52$ eV, respectively, by
inspecting the Ir $L_3$ edge RIXS response in the energy range relevant to
spin-orbital excitations. ii) We show that experimental~\cite{Ohgushi2013}
and theoretical~\cite{Bogdanov2012} results can be reconciled in view of the
nontrivial dependence of the Ir $L_2$/$L_3$ RMXS intensity ratio on the magnetic
moment direction~\cite{MorettiSala2014}. Furthermore, iii) we understand that
the sign of magnetic interactions is unchanged with respect to the ideal
$j_{\mathrm{eff}}=1/2$ case because the even occupancy and the phase relation of
the $yx$ and $zx$ orbitals is preserved. Finally, iv) we clarified how the Mott
insulating state survives in CaIrO$_3$ despite the severe departure from the
$j_{\mathrm{eff}}=1/2$ ground state.

\emph{Acknowledgments -} Enlightening discussions with G. Jackeli are
kindly acknowledged. The experiment largely profited from technical support and
expertise of R. Verbeni and C. Henriquet.

\end{document}